\documentclass[twocolumn]{aastex631}
\usepackage{amsmath}
\usepackage{amssymb}
\usepackage{verbatim}

\newcommand{\alfven}{Alfv\'{e}n~}
\renewcommand{\vec}[1]{\boldsymbol{#1}}

\shorttitle{Turbulence in Weakly-ionized Disks}
\shortauthors{Rea \& Simon}

\begin{document}

\title{Turbulence Can Persist in the Inner Regions of Weakly-Ionized Planet Forming Disks}

\correspondingauthor{David G. Rea}
\email{drea1@iastate.edu}

\author[0000-0002-5000-2747]{David G. Rea}
\affiliation{Department of Physics and Astronomy, Iowa State University, Ames, IA, 50010, USA}

\author[0000-0002-3771-8054]{Jacob B. Simon}
\affiliation{Department of Physics and Astronomy, Iowa State University, Ames, IA, 50010, USA}

\begin{abstract}

Identifying the mechanisms responsible for angular momentum transport in protoplanetary disks, and the extent to which those mechanisms produce turbulence, is a crucial problem in understanding planet formation. The bulk of the gas in protoplanetary disks is weakly ionized, which leads to the emergence of three non-ideal effects, Ohmic diffusion, ambipolar diffusion, and the Hall effect. These low-ionization processes can in some cases suppress turbulence driven by the magnetorotational instability (MRI). However, it has recently been shown that these non-ideal terms can also affect the dynamics of the gas in fundamentally different ways than simple diffusion. In order to further study the role of low-ionization on disk gas dynamics, we carry out a 3D local shearing box simulation with both Ohmic diffusion and ambipolar diffusion and an additional simulation with the Hall effect included. The strength of each non-ideal term, when present, is representative of gas at a radius of 5 AU in a realistic protoplanetary disk. We find the Hall effect increases the saturation strength of the magnetic field, but does not necessarily drive turbulence, consistent with previous work. However, interactions between ambipolar diffusion and the Keplerian shear lead to the ambipolar diffusion shear instability (ADSI), which can drive the initial growth, not damping, of magnetic perturbations. To our knowledge, this is the first work that explicitly demonstrates the viability of the ADSI in the non-linear regime within protoplanetary disks. At later times in the disk, the MRI (reduced in strength by ambipolar-diffusion), may also be present in regions of weak magnetic field between strong concentrations of vertical magnetic flux and sustain turbulence locally in protoplanetary disks.
 
\end{abstract}

\keywords{Circumstellar disks (235), Protoplanetary disks (1300), Magnetohydrodynamical simulations (1966), Planet formation (1241), Accretion (14)}

\section{Introduction} \label{sec:introduction}

The connection between the magnetic field and gas in protoplanetary disks is important in understanding the dynamics and long-term evolution of such disks. Magnetic fields are thought to drive disk accretion by redistributing angular momemtum (i.e., moving some gas to large radii) via laminar and/or turbulent torques on the gas \citep{Lesur+_2014,Balbus&Hawley_1998} or by the vertical and radial outflow of gas entrained in a large-scale wind \citep{Blandford&Payne_1982,Bai_2016,Bethune+_2017,Aoyama&Bai_2023}.

Turbulence was long thought to play the dominant role in driving accretion; turbulence redistributes angular momentum radially via enhanced viscosity between differentially rotating shear layers \citep{Shakura&Sunyaev_1973,Pringle_1981}. However, it is also extremely significant to the planet formation process. Turbulence drives collisions between dust grains \citep{Guttler+_2010,Kothe_2016} potentially disrupting the clumping of mm--cm pebbles \citep{Gole+_2016,Yang+_2018,Umurhan+_2020,Chen&Lin_2020,Lim+_2024}, and, at the other end of the planet formation scale, can generate stochastic torques that affect the migration of protoplanets (e.g. \citealt{Nelson&Papaloizou_2004,Paardekooper+_2011}). For many years, the mechanism thought to be the primary driver of turbulence (and therefore, accretion) in disks was the magnetorotational instability (MRI; \citealt{Balbus&Hawley_1998}). The MRI arises when a sufficiently weak magnetic field (weak usually equates to subthermal in this context) is coupled to gas orbiting with an outwardly decreasing angular velocity gradient (as is the case with Keplerian disks). The MRI is most efficient when the disk gas is well-ionized.

However, other than very close to the star, protoplanetary disks are cold (thus negating thermal ionization) with external sources of ionization (e.g. X-rays) being strongly attenuated in the disk surface; this leads to a poorly ionized midplane region (e.g. \citealt{Gammie_1996,Perez-Becker&Chiang_2011}). In this region, which contains the bulk of disk material, the MRI is heavily modified by collisional interactions between ions, electrons, and neutral species \citep{Balbus&Terquem_2001,Sano&Stone_2002_i,Sano&Stone_2002_ii,Desch_2004,Kunz&Balbus_2004,Wardle&Salmeron_2012}. Ohmic diffusion is thought to be dominant and quench MRI-driven turbulence (e.g. \citealt{Fleming+_2000,Fleming&Stone_2003}) at radii $\lesssim 1$ AU, though the precise radii at which this occurs is sensitive to the disk surface density profile \citep{Wardle_2007} and the Hall effect appears to also be important wherever Ohmic diffusion is strong (e.g. \citealt{Balbus&Terquem_2001,Wardle&Salmeron_2012,Rea+_2024}). At radii $\gtrsim 30$ AU, ambipolar diffusion similarly damps or suppresses the MRI \citep{Simon+_2013_I:weakacc,Simon+_2013_II:strongacc,Cui&Bai_2021}. At intermediate radii the Hall effect is typically the strongest low-ionization process, though ambipolar diffusion can be comparable in strength \citep{Rea+_2024}. The Hall effect can transport angular momentum via persistent magnetic torques on the gas \citep{Lesur+_2014,Bai_2015,Simon+_2015_magneticacc,Rea+_2024} or via intermittent bursts of turbulence \citep{Simon+_2015_magneticacc}, depending on the orientation of the vertical magnetic field \citep{Balbus&Terquem_2001,Kunz_2008} and, in the cases of the turbulent bursts, the strength of ambipolar diffusion \citep{Simon+_2015_magneticacc}.

Although Ohmic diffusion is purely diffusive, the Hall effect and ambipolar diffusion may be able to produce turbulence in their own right. The Hall Shear Instability (HSI; \citealt{Kunz_2008}) is the result of the Hall effect rotating azimuthal magnetic field into the radial magnetic field, which is then sheared into a stronger azimuthal field, completing a positive feedback loop. In this manner small-scale magnetic perturbations can experience exponential growth rates comparable to the MRI. The HSI causes strong laminar azimuthal flows \citep{Lesur+_2014,Rea+_2024}, but it is clear that the HSI can also drive turbulence if ambipolar diffusion is sufficiently weak \citep{Simon+_2015_magneticacc}. Because Keplerian shear leads to the production of azimuthal field from radial, any process that converts azimuthal field to radial field will produce a similar positive feedback loop that leads to growth.  The Ambipolar Diffusion Shear Instability (ADSI; \citealt{Kunz_2008}) occurs because ambipolar diffusion damps magnetic fluctuations oriented parallel to the magnetic field but does not damp perpendicular fluctuations; this leads to radial field remaining such that it can be then sheared back into the azimuthal direction, thus leading to a positive feedback loop with the shear. Magnetic fluctuations grow as a result, provided that ambipolar diffusion is not too strong. Strong ambipolar diffusion will damp or outright quench growth via the ADSI, though the exact threshold where this occurs is sensitive to the strength and geometry of the magnetic field, and the strength of the shear \citep{Desch_2004,Kunz_2008}. At a given location in a real protoplanetary disk, the magnetic field may be subject to a mixture of the MRI with the HSI \citep{Balbus&Terquem_2001,Kunz_2008,Wardle&Salmeron_2012} and the ADSI \citep{Kunz&Balbus_2004,Desch_2004,Kunz_2008}, while also contending with damping from Ohmic and ambipolar diffusion.

Observations of these disks do not yet paint a coherent picture of turbulence. Some systems exhibit a lack of turbulent line broadening (e.g. \citealt{Flaherty+_2017,Flaherty+_2018}) and edge-on disk observations reveal extremely thin dust layers (e.g. \citealt{Doi&Kataoka_2021,Villenave+_2020,Villenave+_2022}), which implies extremely weak turbulence in these systems. However, strong turbulence has been detected in both DM Tau \citep{Flaherty+_2020} and IM Lup \citep{Flaherty+_2024,Paneque-Carreno+_2024}, though this inferred turbulence has only been confirmed for large distances from the star. Turbulence has been detected at the disk surface of the young stellar object SVS 13 \citep{Carr+_2004}, but only in the hot molecular gas ($> 10^3$~K) at distances $\lesssim 0.3$~AU. Thus, the question of turbulence in the inner $\lesssim 30$~au regions of disks remains one that is best addressed by modeling for now.

 \citet{Rea+_2024} found that strong turbulence (gas velocity fluctuations $\gtrsim 5\times 10^{-2}$ the sound speed $c_s$ at the disk midplane, and up to $\sim c_s$ at the disk surface) could still be found at 5 AU in numerical simulations with realistic strengths of Ohmic diffusion, ambipolar diffusion, and the Hall effect. Despite the Hall effect being the dominant non-ideal effect, it contributed mostly to large-scale, laminar flows when the magnetic field was aligned with disk angular momentum vector. The turbulence was insensitive to the orientation of the field, and therefore insensitive to the HSI, though non-axisymmetric modes HSI \citep{Simon+_2015_magneticacc} may have been at work regardless of the orientation of the initial magnetic field.

This paper will be a detailed examination on the physical reason for the generation of turbulence in the terrestrial planet formation region. In Section \ref{sec:methods}, we describe our simulations and disk model. In Section \ref{sec:results}, we examine how the Hall effect and ambipolar diffusion influence our simulations. In Section \ref{sec:linear_regime} we determine which MHD instabilities are present in our simulations, and in Section \ref{sec:nonlinear_regime} we determine if those mechanisms can sustain turbulence in the long-term. In Section \ref{sec:discussion}, we discuss caveats of linear theory and limitations of numerical simulations. In Section \ref{sec:summary}, we summarize our results and discuss the implications for planet formation.

\section{Methods} \label{sec:methods}

We use the Athena code \citep{Stone+_2008} to carry out 3D, local shearing box simulations \citep{Hawley+_1995,Stone&Gardiner_2010} of a patch of a protoplanetary disk. The shearing box has Cartesian coordinates $(x, y, z)$ corresponding to the radial, azimuthal, and vertical dimensions, respectively, and co-orbits the central star with Keplerian angular velocity $\Omega\hat{z}$. We set periodic boundaries in $y$, shear-periodic boundaries in $x$ \citep{Hawley+_1995}, and outflow boundary conditions in the $z$-dimension as in \citet{Simon+_2015_magneticacc}. Athena solves the continuity equation,

\begin{align} \label{eq:continuity}
    \partial_t\rho + \vec{\nabla}\cdot\rho\vec{v} = 0,
\end{align}

\noindent
the momentum equation,

\begin{multline} \label{eq:momentum}
    \partial_t(\rho \vec{v}) + \vec{\nabla}\cdot(\rho\vec{vv} - \vec{BB}) + \vec{\nabla}\left(P + \frac{1}{2}B^2\right) \\
     = 2\rho q\Omega^2\vec{x} - \rho\Omega^2\vec{z} - 2\Omega\vec{\hat{z}}\times\rho\vec{v},
\end{multline}

\noindent
and the induction equation,

\begin{multline} \label{eq:induction}
    \partial_t\vec{B} - \vec{\nabla}\times(\vec{v}\times \vec{B}) \\
    = -\vec\nabla\times\left[ \eta_{\rm O}\vec{J} + \eta_{\rm H}\frac{\vec{J}\times \vec{B}}{B} - \eta_{\rm A}\frac{(\vec{J}\times \vec{B})\times \vec{B}}{B^2} \right].
\end{multline}

\noindent
Here $\rho$ is the gas mass density, $\vec{v}$ is the gas velocity, $\vec{B}$ is the magnetic field which has absorbed a factor of $1/\sqrt{4\pi}$, and $J = \vec{\nabla}\times\vec{B}$ is the current density. From left to right, the source terms of the momentum equation represent radial tidal forces, vertical gravity, and the Coriolis force. We implement an isothermal equation of state with gas pressure $P = \rho c_s^2$, where

\begin{align} \label{eq:cs}
    c_s = \sqrt{\frac{k_{\rm B}T}{\mu m_p}}
\end{align}

\noindent
is the isothermal sound speed, with Boltzmann constant $k_B$, mean molecular weight $\mu = 2.33$, and proton mass $m_p$. The strength of magnetic diffusion is determined by the Ohmic ($\eta_{\rm O}$), Hall ($\eta_{\rm H}$) and ambipolar ($\eta_{\rm A}$) diffusivities \citep{Wardle_2007,Rea+_2024}, and characterized by their respective Elsasser numbers

\begin{align}
    \Lambda &= \frac{v_{\rm A}^2}{\eta_{\rm O}\Omega}, \\
    {\rm Ha} &= \frac{v_{\rm A}^2}{\eta_{\rm H}\Omega}, \\
    {\rm Am} &= \frac{v_{\rm A}^2}{\eta_{\rm A}\Omega},
\end{align}

\noindent
where $v_{\rm A}$ is the magnitude of the \alfven velocity,

\begin{equation}
    \vec{v_{\rm A}} = \frac{\vec{B}}{\sqrt{\rho}}.
\end{equation}

The gas is initially in hydrostatic equilibrium distributed vertically as a Gaussian profile with scale height

\begin{align}
    H = \frac{c_s}{\Omega}
\end{align}

\noindent
and threaded by a vertical magnetic field with $\beta_0~=~+10^4$, where $\beta_0$ is the initial plasma parameter at the disk midplane

\begin{align} \label{eq:beta}
    \beta_0 = \frac{B_{z,0}}{|B_{z,0}|}\frac{2P_0}{|\vec{B}_0|^2}.
\end{align}

\noindent
and the sign of $\beta_0$ indicates the orientation of the initial vertical field $(\vec{\Omega}\cdot\vec{B}_0)$, which is important when the Hall effect is present \citep{Wardle&Ng_1999,Balbus&Terquem_2001,Kunz_2008,Bai_2015,Simon+_2015_magneticacc,Rea+_2024}. We use the disk surface density profile, temperature profile, and vertically stratified ionization prescription --- with contributions from radioactive decay, far-ultraviolet photons, X-rays, and cosmic rays --- of \citet{Rea+_2024}, but with a decreased cosmic ray ionization rate $\zeta_{\rm CR,0} = 10^{-17}~{\rm s^{-1}}$ \citep{Umebayashi&Nakano_1981,Lesur+_2014,Bai_2015}. This creates a highly ionized disk surface, but a poorly ionized disk midplane.

In this work, we focus on two numerical simulations whose main difference is the presence or absence of the Hall effect (Table \ref{table:simulations}). In OAH ${\rm Ha} = 0.054$ initially at the midplane, and in OA ${\rm Ha} = \infty$ (the Hall effect is not present). With these simulations we will isolate the influence of the Hall effect on the behavior of the disk. The remaining Elsasser numbers have initial midplane values of ${\Lambda} = 1.2$ and ${\rm Am} = 0.55$. $\Lambda$ and Ha (when not effectively set to $\infty$) are not static because they are functions of the magnetic field, which grows in strength over time. We employ a spatial resolution of 32 zones per $H$ and a computational domain of size $(L_x,L_y,L_z) = (4H, 6H, 12H)$. This resolution is the same as in \citet{Rea+_2024} which previously measured magnetically driven turbulence, and is comparable to (but often more resolved than) other local simulations exploring non-ideal magnetic effects (e.g. \citealt{Lesur+_2014,Simon+_2015_magneticacc}). The simulations are each run for $1000~\Omega^{-1}$.

\section{Results} \label{sec:results}

\begin{deluxetable*}{lccccccccc}
    \tablecaption{Shearing Box Simulations \label{table:simulations}}
    \colnumbers
    \tablehead{
        \colhead{Name} &
        \colhead{$\zeta_{\rm CR,0}/s^{-1}$} &
        \colhead{$\beta_0$} &
        \colhead{Ha$_0$} &
        \colhead{$\overline{\langle\mathcal{M}_{xy}\rangle}_{\rm mid}$} &
        \colhead{$\overline{\langle\mathcal{M}'_{xy}\rangle}_{\rm mid}$} &
        \colhead{$\overline{\langle\mathcal{R}_{xy}\rangle}_{\rm mid}$} &
        \colhead{$\overline{\langle\delta v/c_s\rangle}_{\rm mid}$} & 
        \colhead{$\overline{\langle\delta v/c_s\rangle}_{\rm mid}^*$} & 
        \colhead{$\sigma/\Omega$}
    }
    \startdata
        OAH & $(-17)$ & $(4)$ & $5.4~(-2)$ & $1.11~(-3)$ & $3.88~(-4)$ & $3.95~(-4)$ & $7.77~(-2)$ & $5.15~(-2)$ & 0.66 \\
        OA & $(-17)$ & $(4)$ & $\infty$ & $3.47~(-4)$ & $1.84~(-4)$ & $1.13~(-4)$ & $4.94~(-2)$ & $4.19~(-2)$ & 0.66 \\
    \enddata
    \tablecomments{Table of parameters and key diagnostics for the shearing box simulations: (1) Simulation name, (2) characteristic cosmic-ray ionization rate at the disk surface, (3) initial plasma $\beta$, (4) initial, midplane Hall Elsasser number, (5) midplane Maxwell stress, (6) small-scale midplane Maxwell stress (Equation \ref{eq:turb_maxwell}), (7) midplane Reynolds stress, (8) midplane turbulent velocity when the magnetic field has even-z symmetry, (9) midplane turbulent velocity when the magnetic field as odd-z symmetry, and (10) the initial growth rate of $B_x$ (Section \ref{sec:linear_regime}). Values written as $a~(b)$ denote $a\times 10^b$. $\langle f\rangle_{\rm mid}$ indicates the vertical spatial average of $\langle f(z)\rangle_{xy}$ (Equation \ref{eq:horizontal_average}) over the innermost $2~H$ of the disk around the midplane; $(2H)^{-1}\int_{-H}^{H} \langle f(z)\rangle_{xy}~dz$.}
\end{deluxetable*}

The Maxwell stress tensor is written as

\begin{equation}
    \mathcal{M}_{ij} = -\frac{B_iB_j}{\rho_0c_s^2}
\end{equation}

\noindent
and the Reynolds stress tensor is

\begin{equation}
    \mathcal{R}_{ij} = \frac{\rho v_iv_j}{\rho_0c_s^2},
\end{equation}

\noindent
where $i,j \in (x,y,z)$ and $v_y$ is the azimuthal component of the velocity with the Keplerian shear already subtracted. The horizontal components $\mathcal{M}_{xy}$ and $\mathcal{R}_{xy}$ are responsible for the radial transport of angular momentum.

\begin{figure*}
    \centering
    \includegraphics[width=\textwidth]{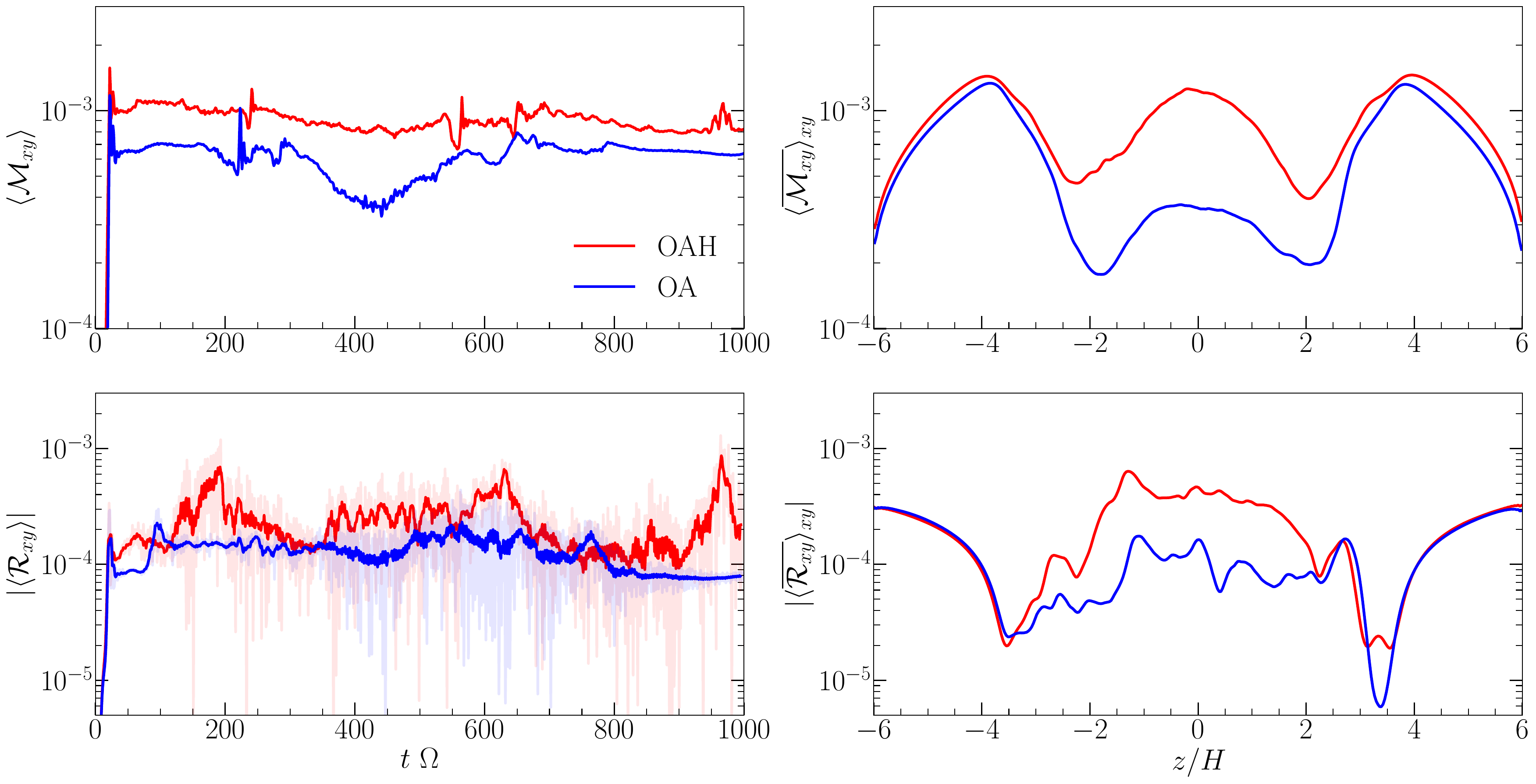}
    \caption{Maxwell (top row) and Reynolds (bottom row) stresses for OA and OAH. The left column shows the box-averaged values as a  function of time, and the right column shows the time- and $xy$-averaged values as a function of height. The Reynolds stress is highly variable; for easier viewing, a solid line which represents a moving average with a window of $\sim 1$ orbit has been added over the real data. The Hall effect increases the stresses by a factor $\sim 3$ in the midplane region.}
    \label{fig:stresses}
\end{figure*}

Figure~\ref{fig:stresses} shows the Maxwell and Reynolds stresses as a function of time (domain-averaged) and as a function of height --- time-averaged from $t = 20~\Omega^{-1}$ until the end of the simulation, where the time average is denoted

\begin{align}
    \overline{f} \equiv \frac{1}{t_1 - t_0}\int_{t_0}^{t_1} f(t)~dt,
\end{align}

\noindent
and horizontally-averaged, where the horizontal spatial average is

\begin{align} \label{eq:horizontal_average}
    \langle f(z)\rangle_{xy} \equiv \frac{1}{L_xL_y}\int f(x,y,z)~dxdy.
\end{align}

\begin{figure*}
    \centering
    \includegraphics[width=\linewidth]{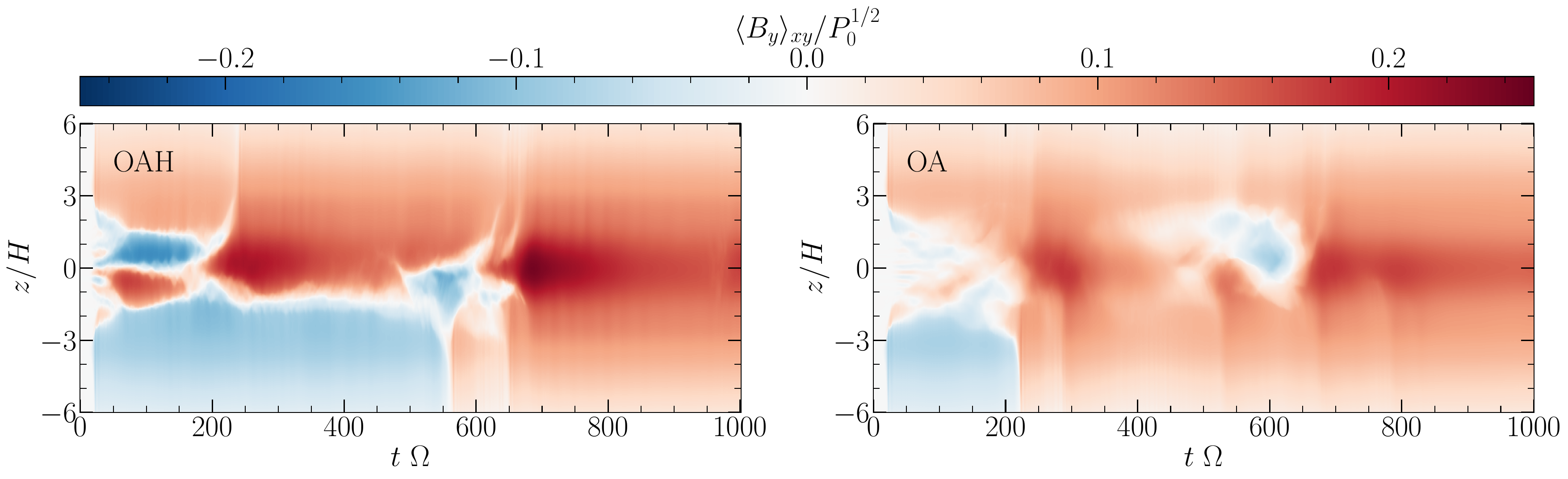}
    \caption{Spacetime diagrams of the azimuthal field $\langle B_y\rangle_{xy}$ for each simulation. Both simulations exhibit multiple transient current sheets up to $200~\Omega^{-1}$ ($\sim~30$ orbits). Thus, we only consider field geometry for $t > 200~\Omega^{-1}$. The field geometry of OAH consists of a long-lived, coherent current sheet that disappears before $700~\Omega^{-1}$. Although two current sheets briefly emerge in OA (from $500\sim 700~\Omega^{-1}$) it does not change the overall odd-z field geometry.}
    \label{fig:by_st}
\end{figure*}

OAH consistently exhibits a stronger Maxwell stress than OA throughout the simulation. Long-term variations are correlated with and thus may be caused by the presence (or absence) of large-scale, long-lived magnetic nulls (current sheets) while short term fluctuations may indicate turbulence.
Such current sheets can be seen in Figure \ref{fig:by_st}, which shows the spacetime diagram (time and horizontal spatial average) of the azimuthal magnetic field $B_y$. OAH exhibits an azimuthal field geometry with approximately $B_y(z) \sim -B_y(-z)$ for a significant portion of its duration, which manifests as a long-lived current sheet near the midplane. While this expression is only true when the current sheet is exactly at the midplane (which is often not the case in numerical simulations), we still refer to this as ``even-z" symmetry based on the poloidal field configuration with which it corresponds \citep{Bai&Stone_2013_wind}. Similarly, OA exhibits a field geometry with $B_y(z) \sim B_y(-z)$, which we refer to as the``odd-z," for most of its duration (though two current sheets  separate and converge between 400--700~$\Omega^{-1}$). These ``even-z" and ``odd-z" designations also allow us to make contact with the literature, such as \citet{Bai&Stone_2013_wind}.

\begin{figure}
    \centering
    \includegraphics[width=\linewidth]{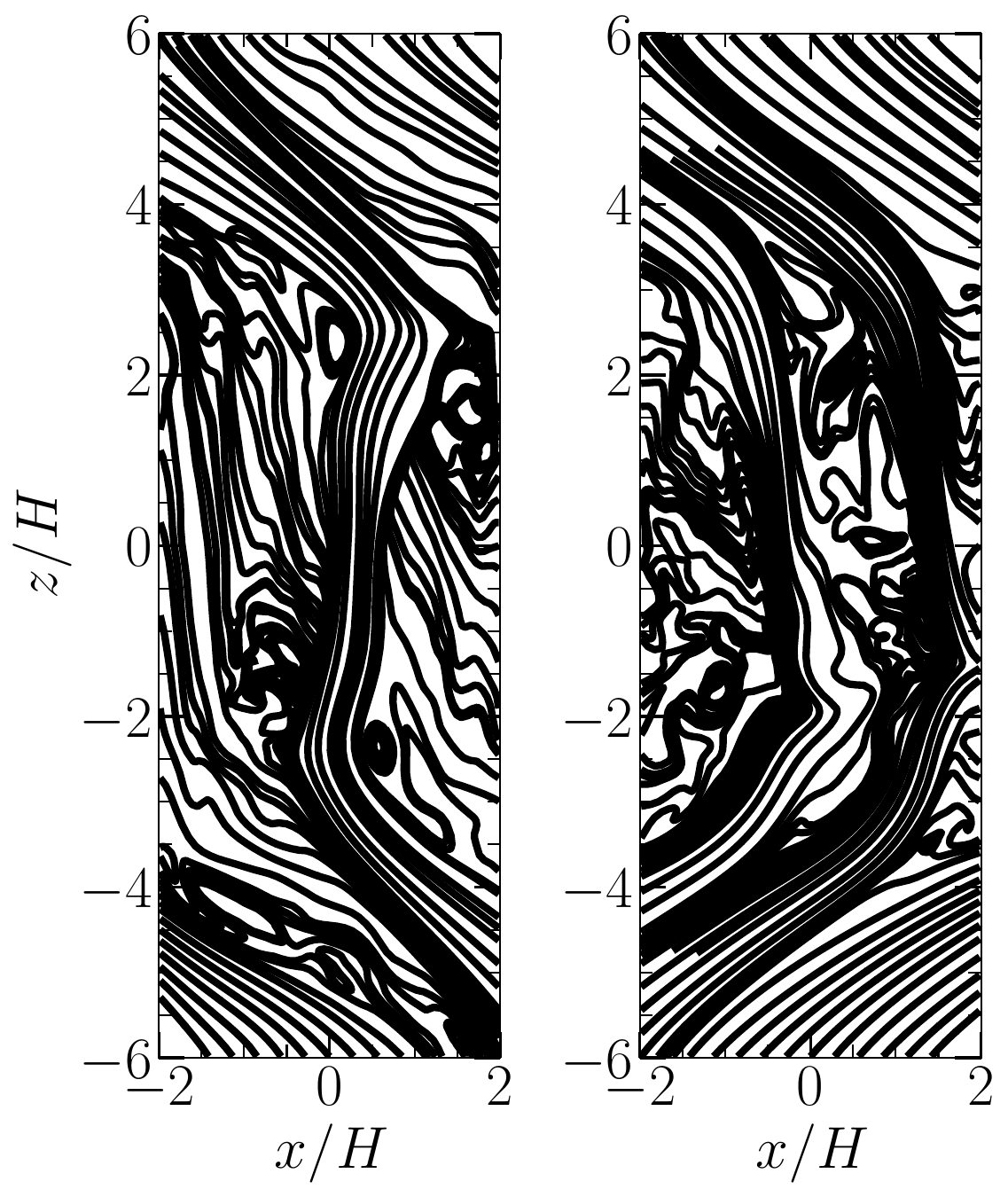}
    \caption{Slice of the poloidal magnetic field lines for OA (left) and OAH (right) at $t = 400~\Omega^{-1}$. A strong radial field at $|z| \gtrsim 4~H$ causes strong Maxwell stress in those regions. In the bulk of the gas, the vertical component dominates the poloidal field, though as a whole, the magnetic field is primarily toroidal. At this time, OA exhibits odd-z symmetry and OAH exhibits even-z symmetry.}
    \label{fig:bfield_structure}
\end{figure}

Returning to Figure \ref{fig:stresses}: The Reynolds stress in OAH and OA appears to share a similar baseline of $\sim 10^{-4}$ throughout the simulation. While $\mathcal{R}_{xy}$ stays within a factor of $\sim2$ of this baseline, OAH possesses large-amplitude fluctuations up to $\mathcal{R}_{xy} \sim 10^{-3}$, an order of magnitude larger. These large-amplitude fluctuations are correlated with changes to the large-scale field geometry, i.e. when the magnetic field transitions from a geometry with a large-scale, long-lived magnetic null, to a geometry without such a null, or vice-versa (with the exception of the large-amplitude fluctuation at $t\sim 950~\Omega^{-1}$ in OAH; this is correlated with a transient current loop that does not manifest in $\langle B_y\rangle_{xy}$ in Figure \ref{fig:by_st}).
The strength of the Maxwell stress greatly varies with distance from the disk midplane due to the geometry of the magnetic field (Figure \ref{fig:bfield_structure}). At the midplane, $\mathcal{M}_{xy}$ is dominated by the azimuthal field $B_y$, which weakens with increasing height. At $z \approx \pm 2~H$ the density has decreased enough for $\beta$ to be approximately unity, and the large-scale field bends radially into $B_x$ (e.g. \citealt{Bai&Stone_2013_wind,Rea+_2024}) though the field is still primarily toroidal. The magnetic field slightly straightens toward the vertical past $|z| \sim 4~H$, where the gas dynamics become dominated by strong outflows, such that the radial field is diverted into vertical field; the value of $\mathcal{M}_{xy}$ decreases accordingly. The transition into an outflow region is also apparent in the Reynolds stress around $\pm 4~H$, as midplane fluctuations give way to a more laminar region near the disk surface. Despite the similarities, we refrain from labeling this outflow as a disk wind due to limitations of the shearing box in capturing global-scale dynamics.

\begin{figure}
    \centering
    \includegraphics[width=\linewidth]{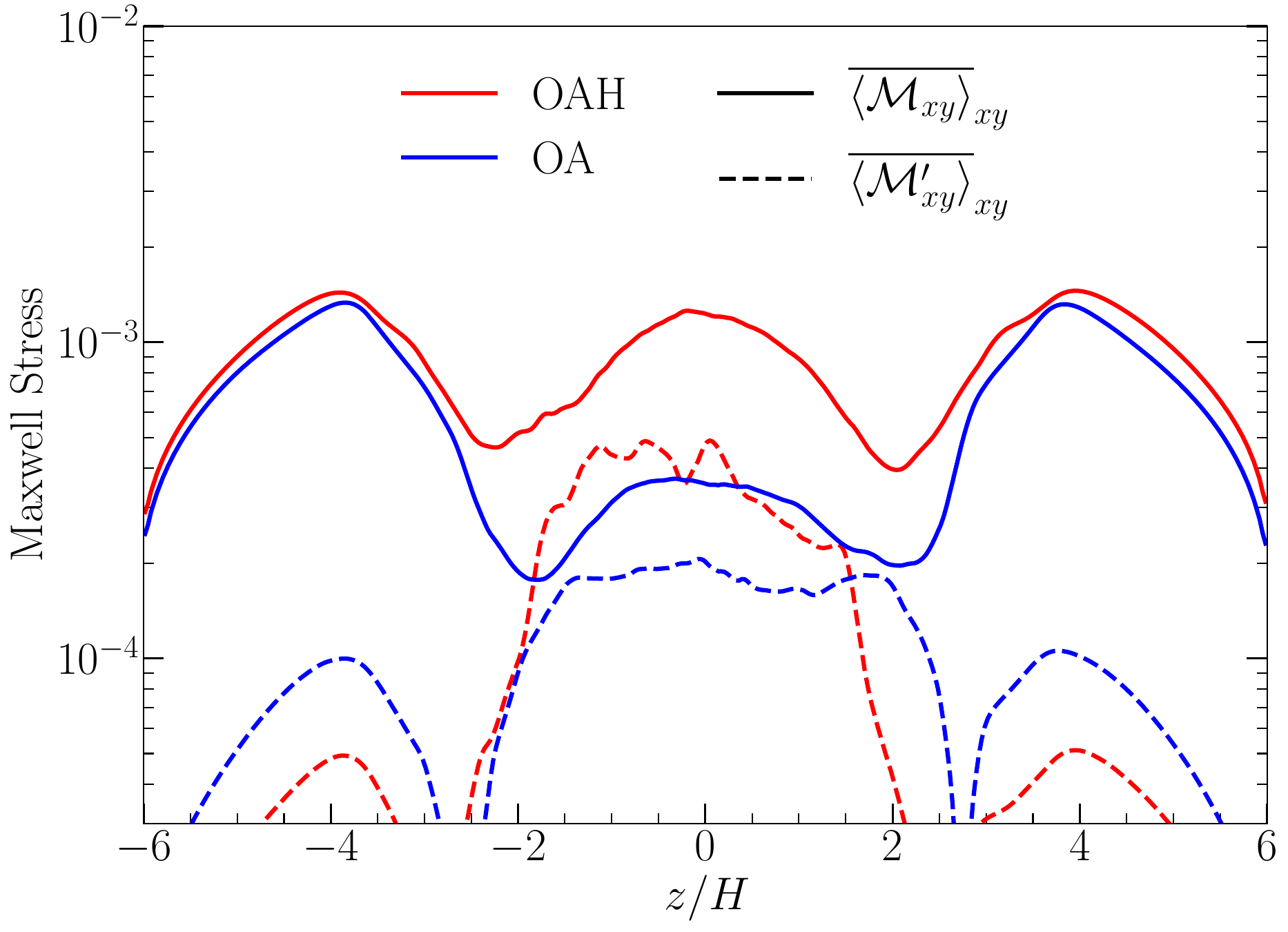}
    \caption{Total $\mathcal{M}_{xy}$ and small-scale $\mathcal{M}'_{xy}$ component of the Maxwell stress. For heights $|z| > 2~H$, the Maxwell stress is almost entirely due to large-scale magnetic fields. In the midplane, the small-scale and large-scale fields contribute equally to the Maxwell stress ($\mathcal{M}'_{xy} \approx \frac{1}{2}\mathcal{M}_{xy}$) regardless of the strength of the Hall effect.}
    \label{fig:maxwell_components}
\end{figure}

Figure \ref{fig:maxwell_components} provides a closer look at the vertical structure of the Maxwell stress. The Hall effect enhances the total Maxwell stress by about a factor of 3 on average in the disk midplane.  This increase is almost entirely due to a much stronger radial magnetic field in the midplane when the Hall effect is present. In the disk surface where the ionization is stronger and ${\rm Ha} \gg 1$ for both simulations, the strengths of the radial field are comparable; therefore the Maxwell stresses are also similar.\footnote{Even though the azimuthal field dominates the Maxwell stress, the difference in the azimuthal field component $\langle B_y\rangle_{xy}$ is only 10--20\% different between the two simulations.}
The small-scale Maxwell stress can be computed by removing the influence of volume-averaged field components, which for simplicity we will refer to as the large-scale Maxwell stress,

\begin{align} \label{eq:turb_maxwell}
    \mathcal{M}'_{xy} = \mathcal{M}_{xy} + \frac{\langle B_x\rangle\langle B_y\rangle}{\rho_0c_s^2}.
\end{align}

\noindent
The Maxwell stress is almost entirely large-scale for heights $|z| \gtrsim 2~H$. However, in the midplane region $\mathcal{M}'_{xy} = \frac{1}{2}\mathcal{M}_{xy}$, i.e. the small-scale and large-scale fields contribute more or less equally to the total Maxwell stress, regardless of whether or not the Hall effect is present.

\begin{figure}
    \centering
    \includegraphics[width=\linewidth]{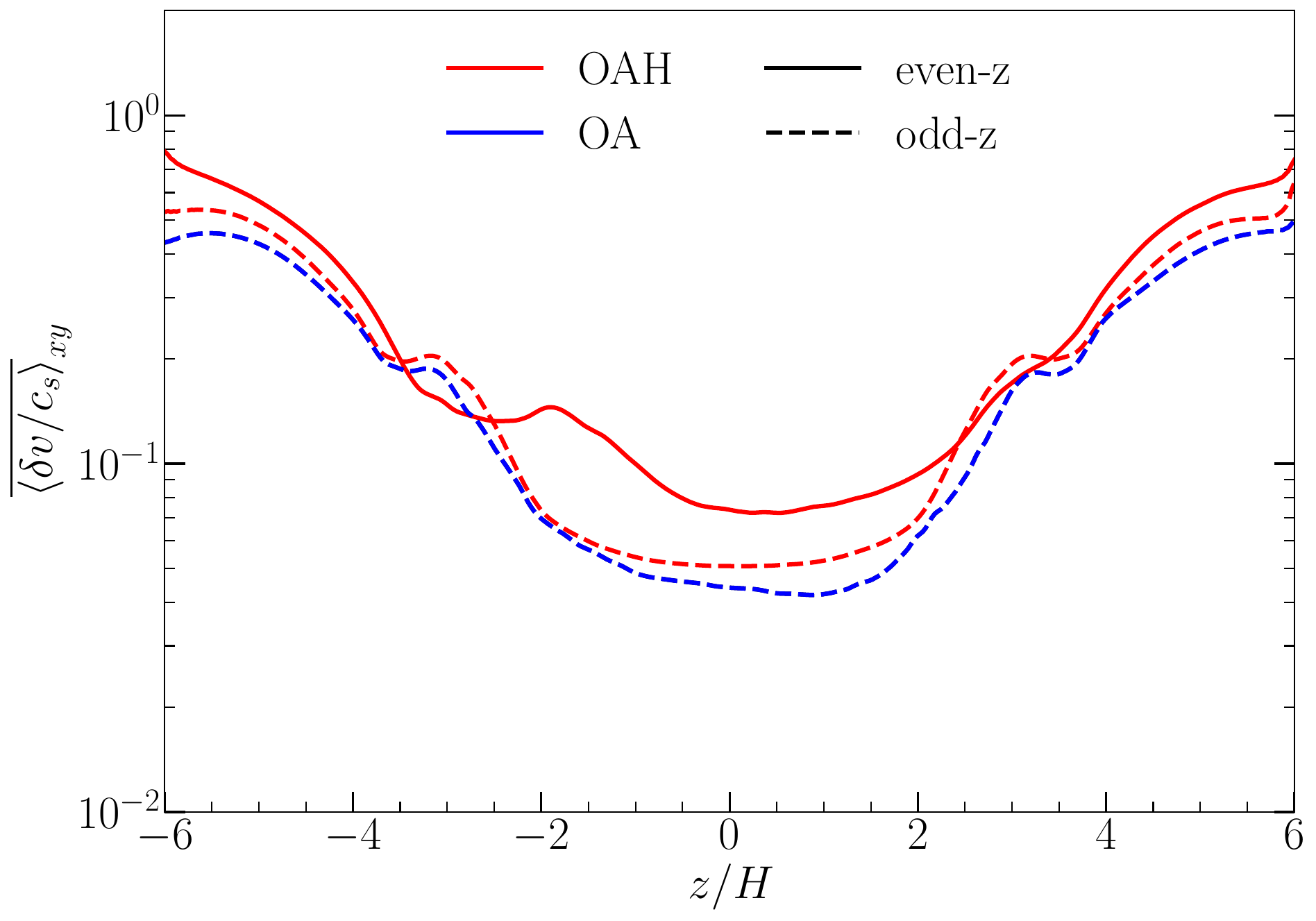}
    \caption{Turbulent velocities in OA and OAH. The velocities are averaged over times with similar azimuthal magnetic field geometries: in OAH, from $200-550~\Omega^{-1}$ and from $600~\Omega^{-1}-1000~\Omega^{-1}$, and in OA, from $200~\Omega^{-1}$ until the end of the simulation.}
    \label{fig:deltav_z}
\end{figure}

The turbulent velocities of a given snapshot are computed by removing the shear and bulk horizontal and zonal flows. With these flows subtracted, the turbulent velocity magnitude is given by

\begin{align}
    \delta v \equiv \sum_{i=x,y,z}\langle\delta v_i^2\rangle^{1/2}
\end{align}

\noindent
where $\delta v_i$ is the velocity fluctuation in a direction $i$, following the method of \citet{Simon+_2015_signatures,Rea+_2024}. Figure \ref{fig:deltav_z} shows the turbulent velocities of OAH and OA. Both simulations have turbulent velocities of $\sim 5\times10^{-2}~c_s$ in the midplane, approaching $\sim c_s$ at heights $\gtrsim~5~H$.
When OAH exhibits even-z geometry which exhibits a long-lived current sheet near the midplane (recall Figure \ref{fig:by_st}) there is slightly stronger turbulence than times with odd-z field geometry. For this reason, we separately time average the turbulent velocities in OAH over times with similar overall field geometry: from $200-550~\Omega^{-1}$ and from $600~\Omega^{-1}-1000~\Omega^{-1}$. The first $200~\Omega^{-1}$ are not included in the turbulent velocity average in order to exclude any contributions from initial transients. OA exhibits an odd-z field geometry for most of its duration, which is not expected to enhance turbulence \citep{Rea+_2024}. Although two current sheets briefly emerge in OA before reconnecting, they do not change the overall azimuthal field geometry and do not significantly enhance the turbulent gas velocities.

These initial diagnostics indicate that non-ideal MHD effects can have significant influence on the structure of protoplanetary disks. In particular, the Hall effect can increase the overall strength of the magnetic field and Maxwell stress. However, the Hall effect does not appear to have a significant impact on the gas turbulence. Instead, a level of turbulence is present in the disk midplane regardless of the presence of the Hall effect, and can be enhanced by changes to the large-scale magnetic field geometry such as magnetic reconnection.

\section{Linear Regime} \label{sec:linear_regime}

\begin{figure}
    \centering
    \includegraphics[width=\linewidth]{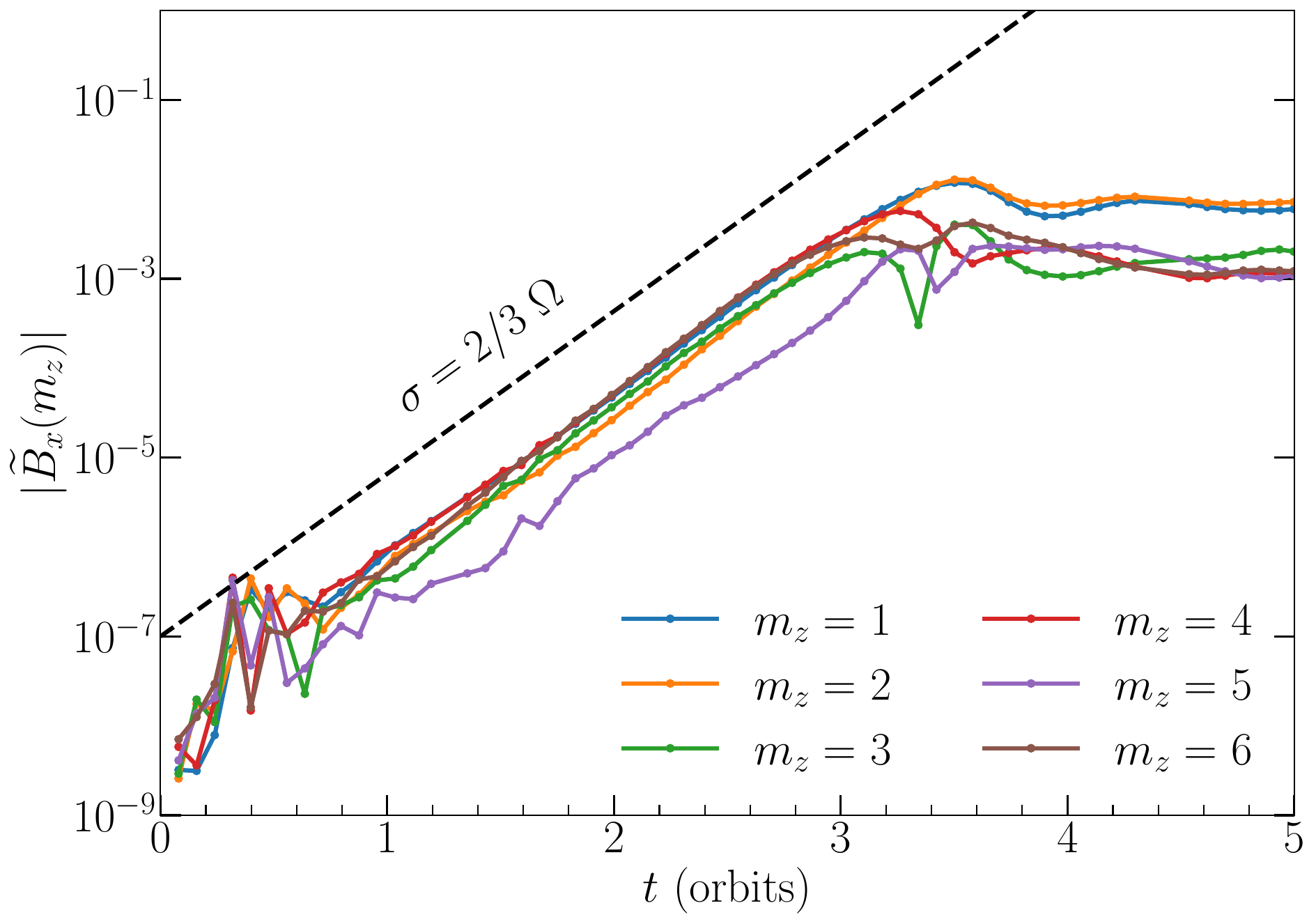}
    \caption{Initial growth of the radial magnetic field in OA for $k_x = k_y = 0$. The first six vertical modes $m_z$ (excluding $m_z = 0$) are shown, where $2\pi m_z = k_zL_z$. The growth rate is extremely fast, being approximately $0.66~\Omega$, and the magnetic field saturates within the first 4 orbits.}
    \label{fig:bx_growth}
\end{figure}

The magnetic field is initially a constant vertical field $\vec{B}_0 = (0, 0, B_{z,0})$ but small fluctuations quickly develop into a pronounced vertical structure. Between $\sim 0.5$ and $\sim 3.5$ orbits, the growth of the radial magnetic field is exponential, with

\begin{align}
    B_x(t) \propto e^{\sigma t}
\end{align}

\noindent
with a growth rate $\sigma \approx0.66~\Omega$. Figure \ref{fig:bx_growth} shows this growth for the first few vertical wavenmubers in OA. Such a setup and subsequent growth can clearly be connected to linear analysis, which considers small-amplitude perturbations to the magnetic field with the form $\exp(\sigma t - i\vec{k}\cdot\vec{x})$, where $\vec{k}$ is the angular wavevector defined by the Fourier transform

\begin{align}
   \widetilde{f}(\vec{k}) = \int f(\vec{x})e^{-i\vec{k}\cdot\vec{x}}~d\vec{x}.
\end{align}

\noindent
The behavior of the perturbations under the influence of the equations of motion (Equations \ref{eq:continuity}--\ref{eq:induction}) is described by a dispersion relation which relates $\sigma$ and $\vec{k}$.

Turbulence is apparent with or without the Hall effect (in OA and OAH; see above), and the growth of the radial magnetic field of OAH is practically identical to that of OA in Figure \ref{fig:bx_growth}. Thus, whatever is driving the turbulence is likely tied to the instability with the linear growth rate shown here. In order to determine the origin of the turbulence (and keep our analysis straightforward), we first focus on simulation OA, which does not include the Hall effect.

\subsection{Expectations from Linear Theory} \label{sec:linear_regime:expectations}

We follow \citet{Desch_2004} to obtain the dispersion relation that contains the influence of Ohmic and ambipolar diffusion:

\begin{equation} \label{eq:disp_relation}
    \sigma^4 + c_3\sigma^3 + c_2\sigma^2 + c_1\sigma + c_0 = 0
\end{equation}

\noindent
with coefficients

\begin{align}
    c_3 &= \omega_{\rm OD} + \omega_{\rm AD} + \omega_T \\
    c_2 &= \kappa^2g_\theta + 2\omega_{\rm A}^2g_l \nonumber\\
    &\quad +\; (\omega_{\rm OD} + \omega_{\rm AD})\omega_T + \omega_{\rm AD}\left(\frac{d\Omega}{d\ln R}\right)g\\
    c_1 &= (\kappa^2g_\theta + \omega_{\rm A}^2g_l)c_3 \\
    c_0 &= \left(\omega_{\rm A}^2g_l + \frac{d\Omega^2}{d\ln R}g_\theta\right)\omega_{\rm A}^2g_l \nonumber\\
    &\quad +\; \kappa^2g_\theta(\omega_{\rm OD} + \omega_{\rm AD})\omega_T \nonumber\\
    &\quad +\; \omega_{\rm AD}\left(\frac{d\Omega}{d\ln R}\right)(\kappa^2g_\theta + \omega_{\rm A}^2g_l)g.
\end{align}

\noindent
where $\kappa$ is the epicyclic frequency, $g_\theta \equiv (\vec{\hat{e}}_k\cdot\vec{\hat{e}}_z)^2$, $g_l \equiv (\vec{\hat{e}}_k\cdot\vec{\hat{e}}_B)^2$, and $g \equiv (\vec{\hat{e}}_k\times\vec{\hat{e}}_B\cdot\vec{\hat{e}}_\phi)(\vec{\hat{e}}_B\cdot\vec{\hat{e}}_\phi)(\vec{\hat{e}}_z\cdot\vec{\hat{e}}_z)$ are geometric factors, and $\omega_{\rm A} = kv_{\rm A}$, $\omega_{\rm OD} = k^2\eta_{\rm O}$, $\omega_{\rm AD} = k^2\eta_{\rm A}$, and $\omega_T = \omega_{\rm OD} + \omega_{\rm AD}g_l$ describe the strength of the nonideal effects. The disk is unstable when $c_0 < 0$.

\begin{figure*}
    \centering
    \includegraphics[width=\textwidth]{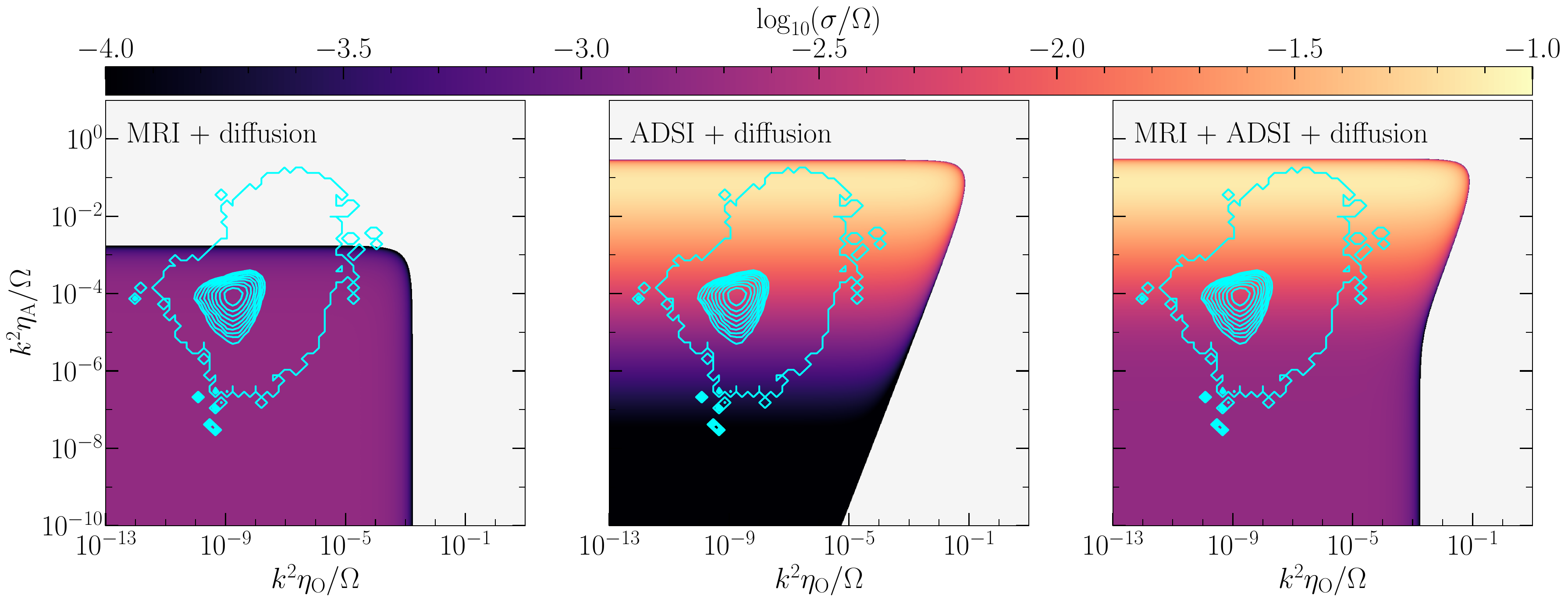}
    \caption{Growth rate of different instabilities at for different combinations of $k^2\eta_{\rm O}$ and $k^2\eta_{\rm A}$ informed by \citet{Desch_2004}, and OA simulation data at $t = 2~\Omega^{-1}$ (blue contours; each contour contains a percentage of the data in increments of 10\%.) To compute growth rates we estimate $\omega_{\rm A} = 10^{-3}$ except in the middle panel, where $\omega_{\rm A} = 0$ (see text for details). Grey regions are stable to both the MRI and ADSI. Ambipolar diffusion is crucial in destabilizing the disk via the ADSI \citep{Kunz_2008}.}
    \label{fig:wod_wad_tmp}
\end{figure*}

This dispersion relation describes a mixture of both the MRI and the ADSI (and diffusion). For example, the first set of two terms in $c_0$ describes \alfven waves and the destabilizing influence from Keplerian rotation which are inherent to the MRI, the second set of two terms describe the stabilizing influence of Ohmic and ambipolar diffusion, and the final two terms describe the destabilizing influence of ambipolar diffusion coupled to the shear.  When $\eta_{\rm O} = \eta_{\rm A} = 0$ (and the magnetic field and perturbations are aligned with the vertical), the dispersion relation resembles that of the ideal MRI of \citet{Balbus&Hawley_1998}. Figure \ref{fig:wod_wad_tmp} shows the resulting growth rates for only the MRI (left), only the ADSI (middle), and both instabilities combined (right), for a given diffusion strength. Cyan contours in each panel map where the data from OA lie in the $k^2\eta_{\rm O}$-$k^2\eta_{\rm A}$ plane only $2~\Omega^{-1}$ after starting the simulation. To extract the growth rates of MRI and ADSI separately from the dispersion relation above, certain assumptions can be made. To extract the behavior of the MRI in the presence of Ohmic and ambipolar diffusion, we artificially set only the ambipolar diffusion-shear coupling term $\omega_{\rm AD}(d\Omega/d\ln R) = 0$ to remove the destabilizing influence of the ADSI, but in general $\omega_{\rm AD} \ne 0$ and $\omega_{\rm OD} \ne 0$. We estimate $\omega_{\rm A} = 10^{-3}$, which roughly corresponds to $kv_{\rm A}$ for fluctuations close to the grid scale in OA. The resulting growth rates are shown in the left panel of Figure \ref{fig:wod_wad_tmp}; we find that as Ohmic or ambipolar diffusion becomes too strong ($k^2\eta \gtrsim kv_{\rm A}$) the disk is stable to the MRI, with stable regions colored grey.  The ability of ambipolar diffusion to damp the MRI is modulated by the geometric factor $g_l$; perturbations perpendicular to the magnetic field ($g_l = 0$) are completely unaffected by ambipolar diffusion. In addition to the strength of magnetic diffusion, the spatial scale is also important: small scales (large $k$) are more readily subject to diffusion, and for a given (nonzero) $\eta_{\rm_O}$ or $\eta_{\rm A}$ there can always be a spatial scale large enough that the disk is unstable to the MRI (\citealt{Wardle_1999,Kunz&Balbus_2004,Desch_2004}; though such a large scale may not be realistic to consider in protoplanetary disks).

To extract the behavior of the ADSI from the MRI under diffusion from the dispersion relation we assume the magnetic field is extremely weak, or $kv_{\rm A}/\Omega \ll 1$.  Under this assumption the magnetic tension force that couples two fluid elements, the heart of the MRI, is damped by magnetic diffusion faster than the tension can be transmitted by \alfven waves. Note that this approximation does not necessarily hold for any part of our simulations; it serves only to remove the terms of the dispersion relation that are responsible for the MRI so that the remaining system can be better understood (see \citealt{Desch_2004} for a similar analysis on the effects of ambipolar diffusion on the MRI). The instability that remains is analogous to the ADSI of \citet{Kunz_2008}, though that work only considered the case of planar shear and no Ohmic diffusion. The middle panel of Figure \ref{fig:wod_wad_tmp} displays a drastically expanded region of the $k^2\eta$-plane where the disk is unstable. If ambipolar diffusion is weak, the growth rate of the ADSI is small. However, at its strongest, the ADSI drives growth rates roughly 40 times larger than the  maximum growth rate (reduced due to ambipolar diffusion) of the MRI. The ADSI-driven growth occurs at values of $k^2\eta_{\rm A}$ where the disk would otherwise be stable to the MRI alone.  As ambipolar diffusion becomes even stronger (or smaller spatial scales are considered), the damping effects of ambipolar diffusion become dominant and the disk is stabilized.

Finally, the right panel of Figure \ref{fig:wod_wad_tmp} shows the growth rates of the full MRI + ADSI + diffusion mixture. Having elucidated the behavior of the separated instabilities, we now compare the actual simulation data of OA (cyan contours) to linear theory. Despite linear analysis predicting that the majority of the simulation is unstable to the MRI, the maximum growth rate of the MRI is only predicted to be $\sim 2\times 10^{-3}~\Omega$. In contrast, a small fraction of the simulation is predicted to have a much larger growth rate of $\sim 7\times 10^{-2}~\Omega$ at values of $k^2\eta_{\rm A}$ for which the MRI is quenched. The exact instability threshold is sensitive to the magnetic field geometry; additionally, the exact values of the growth rates may be different from the actual growth rates present in simulations due to certain assumptions of linear analysis (see e.g.\citealt{Desch_2004,Kunz_2008}) especially as the background magnetic field continues to grow. However, we do not expect the relative growth rate amplitudes between the MRI and ADSI to be significantly changed, especially at early times when the simulation closely resembles the setup considered by linear analyses.

\subsection{Analysis of Simulation Linear Growth Rates} \label{sec:linear_regime:analysis}

Having examined the initial instability in OA, we now consider what linear analysis predicts for the growth rate between $\sim 0.5$ and $\sim 3.5$ orbits, when the simulation undergoes clear and consistent exponential growth (Figure \ref{fig:bx_growth}). We examine a single snapshot of OA and OAH at $10~\Omega^{-1}$ ($\sim 1.6$ orbits), as the growth rate is constant from $0.5-3.5$ orbits.

\begin{figure*}
    \centering
    \includegraphics[width=\textwidth]{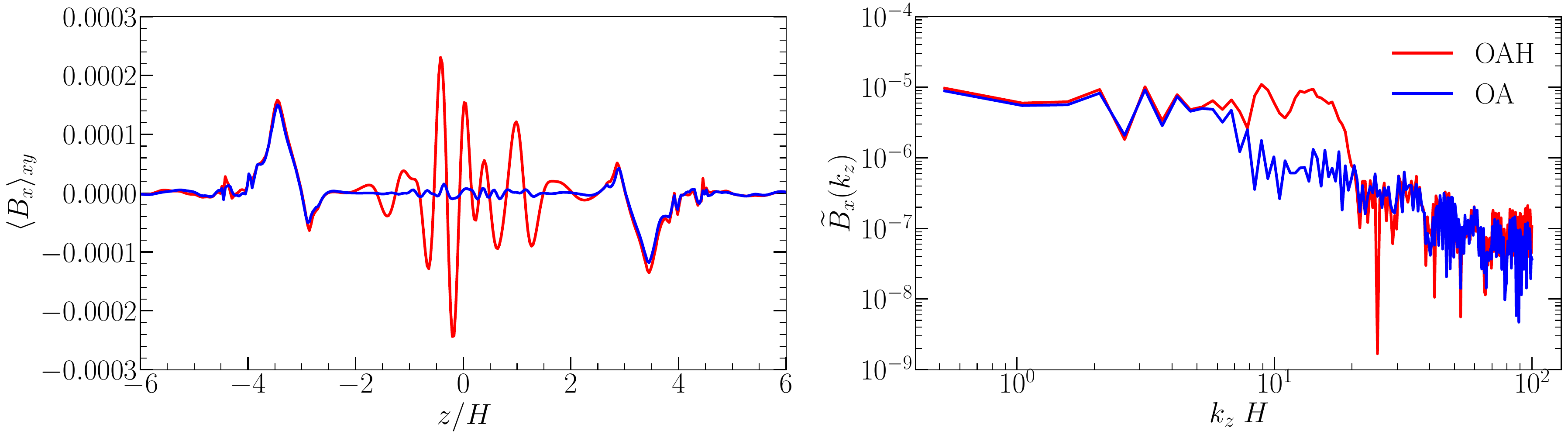}
    \caption{Vertical structure of OA and OAH at $t = 10~\Omega^{-1}$. (Left) Horizontally averaged radial field. Both runs exhibit fast growth at $\sim\pm 3.5~H$. The Hall effect causes relatively large-amplitude small-scale structure within $2~H$ of the midplane in OAH. (Right) Fourier transform of the radial field at $k_x=k_y=0$. The Hall-induced modes are present for $8 < k_zH < 20$.}
    \label{fig:initial_vertical_structure}
\end{figure*}

\begin{figure*}
    \centering
    \includegraphics[width=\linewidth]{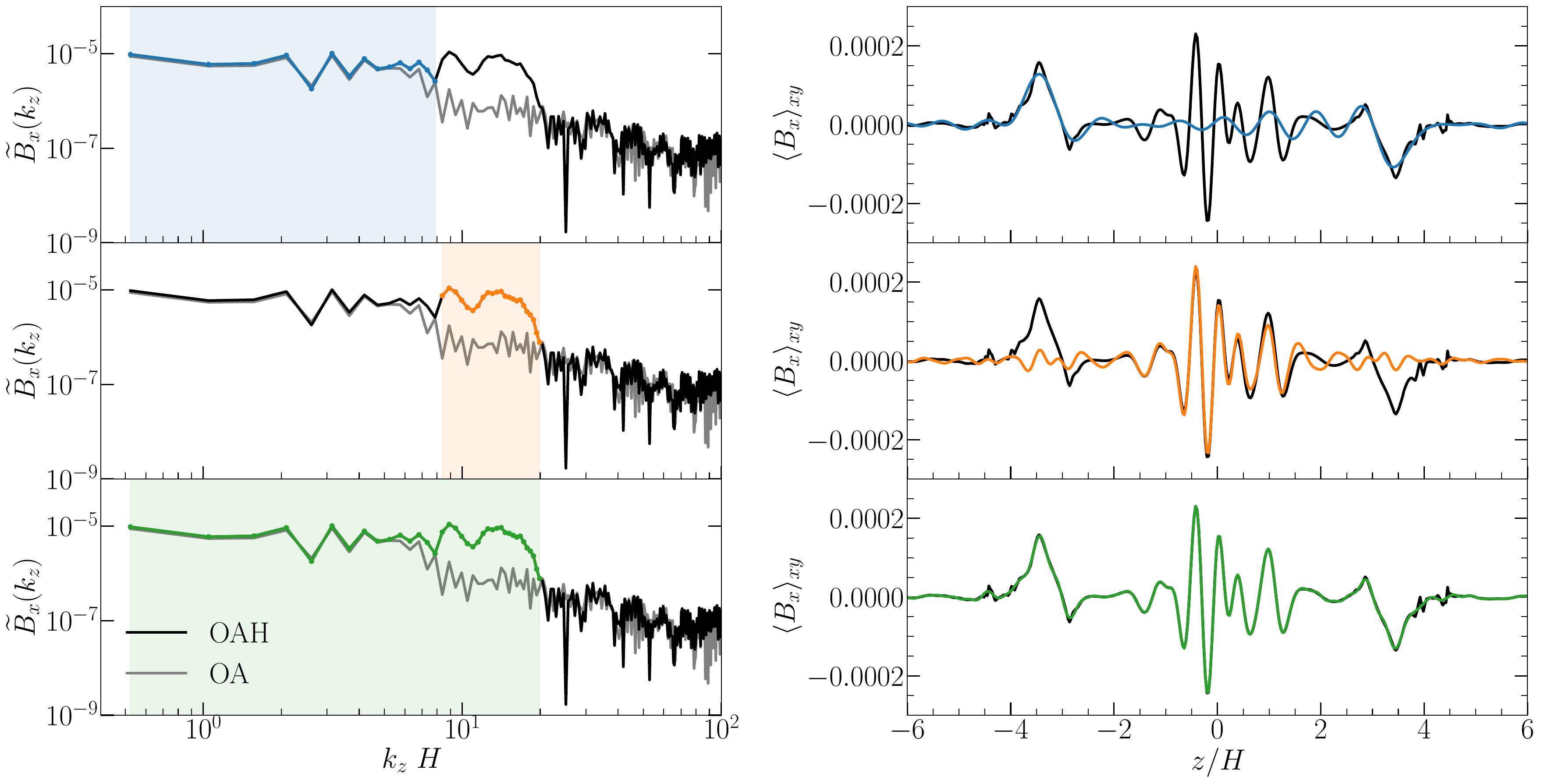}
    \caption{Left column: Vertical structure of $\widetilde{B}_x$ in Fourier space. Right column: vertical structure of $B_x$ in physical space. The physical structure at $\pm 3.5~H$ is best reconstructed by Fourier modes with $k_z < 8 ~H^{-1}$ or wavelengths $L \gtrsim H$ (top panels). The small-scale physical structure in the midplane, which is only apparent when the Hall effect is present, is best reconstructed by Fourier modes between $8 < k_z H < 20$, or wavelengths $0.3 < L/H < 1$ (middle panel). The vertical structure at all scales is well represented by wavelengths $L > 0.3~H$ (bottom panel). The scales at which initial vertical structures arise are directly related to features in Fourier space.}
    \label{fig:fft_space_comp}
\end{figure*}

Figure \ref{fig:initial_vertical_structure} shows the vertical structure of $B_x$ in OA and OAH after $10~\Omega^{-1}$. The left panel shows the horizontally averaged radial field for each simulation as a function of height, while the right panel expresses this information in terms of spatial scales $k_z$ and the corresponding Fourier transform $\widetilde{B}_x(0,0,k_z)$.  Both OA and OAH exhibit near identical features of (relatively) large amplitude local growth at $z\sim\pm3.5~H$, while OAH additionally exhibits smaller scale structure in the midplane region $|z| \lesssim 2~H$. These spatial structures clearly correspond to features in Fourier space. We are able to spatially reconstruct these features with distinct sets of Fourier modes, as shown in Figure \ref{fig:fft_space_comp}. The $\pm 3.5~H$ growth is well reconstructed by vertical modes with $k_z < 8~H^{-1}$, or wavelengths $L \gtrsim H$. The midplane growth of the magnetic field is composed of shorter wavelengths, with $k_z$ approximately between 8 and $20~H^{-1}$. Combined, these modes contain most of the power and are the most important for reconstructing the features of $B_x$.

We find that this growth most closely matches the growth rate of the mixture of the MRI and ADSI when Ohmic diffusion is neglected and only vertical perturbations are considered (similar to the linear setup examined by \citealt{Kunz&Balbus_2004}). To determine this, we use the values of $\vec{k}$, $\vec{\widetilde{v}_{\rm A}}(\vec{k})$ and Am from the OA simulation to compute the growth rate of the magnetic field. We use a single representative value of ${\rm Am} = 3$ to describe the strength of ambipolar diffusion at $\pm 3.5~H$, the location of the strongest growth. Using these values we solve the quartic dispersion relation exactly for unstable (positive, real) values of $\sigma(\vec{k})$. The maximum growth rate of the MRI and ADSI mixture computed in this manner is predicted to be $0.60~\Omega$, which is in excellent agreement with the actual growth rate of $0.66~\Omega$ exhibited in the simulations.

The HSI \citep{Kunz_2008} is a strong candidate for producing the magnetic field midplane modes in OAH between $-2 < z/H < 2$, in particular because of the large scale structure and strong azimuthal field similarities between this work and \citet{Rea+_2024} and \citet{Lesur+_2014}. We compare the linear theory of \citet{Kunz_2008} using the same procedure detailed above for OAH. The Hall-induced modes are restricted to within $\pm 2~H$ of the midplane, where the strength of the Hall effect is only weakly sensitive to height; therefore we assume ${\rm Ha} = \langle {\rm Ha}\rangle_\mathcal{G} = 0.08$ in this region, where 

\begin{align}
    \langle f \rangle_{\mathcal{G}} \equiv \exp\left(\frac{\int_\mathcal{V} \ln f(\vec{x})~d\vec{x}}{\int_\mathcal{V}d\vec{x}}\right)
\end{align}

\noindent
is the geometric average. This method predicts that the HSI should have a growth rate of $0.36~\Omega$, which is significantly smaller than the real growth rate of $0.66~\Omega$ (though still quite fast, being only a factor of $\sim 2$ slower than the maximum growth rate of the ideal MRI). That both the OA and OAH simulations have identical growth rates in the linear regime suggests that the primary mechanism contributing to turbulence in the nonlinear state is the ADSI and MRI, while the Hall effect acts as a secondary instability with smaller growth rate. This is consistent with \citet{Rea+_2024}, who found that the strength of gas turbulence was insensitive to the orientation of the magnetic field --- and therefore, the Hall effect and the HSI --- despite the Hall effect being important for laminar gas flows and field structures.

\section{Nonlinear Regime} \label{sec:nonlinear_regime}

Linear theory is a good predictor of the mechanisms responsible for the initial growth of individual modes, and while it remains a useful tool, it does not provide complete information on the behavior of the disk at later times when the system becomes fully nonlinear. At later times, the magnetic field is much stronger and the MRI is expected to be more prominent as $kv_{\rm A} > k^2\eta_{\rm O,A}$. In the nonlinear regime, the MRI is suppressed in environments with strong Ohmic \citep{Gammie_1996,Fleming+_2000} and/or ambipolar diffusion \citep{Bai&Stone_2011,Simon+_2013_II:strongacc,Simon+_2013_I:weakacc,Cui&Bai_2021}, though the MRI is not completely quenched by ambipolar diffusion in the presence of a weak net vertical magnetic field (\citealt{Simon+_2013_II:strongacc}; it is possible that this is a manifestation of the ADSI in that work). Linear analysis shows that arbitrarily strong ambipolar diffusion can destabilize the disk at different spatial scales, provided that the magnetic field is sufficiently weak \citep{Desch_2004,Kunz&Balbus_2004}. Specifically, unstable modes will exist for wavelengths greater than the critical wavelength \citep{Wardle_1999,Bai&Stone_2011}

\begin{align} \label{eq:crit_wavelength}
    \frac{L_c}{H} = 5.13\left(1 + \frac{1}{\rm Am^2}\right)^{1/2}\beta^{-1/2}
\end{align}

This condition was used by \citet{Bai&Stone_2011} to ensure that the most unstable mode of the MRI was contained within their simulation domain for the initial chosen values of Am and $\beta_0$. When $L_c = H$, it is similar to the empirically discovered Am-$\beta$ threshold for the MRI of \citet{Bai&Stone_2011} which found that the MRI was permitted for a wide range of Am, provided that $\beta$ was sufficiently weak. However, \citet{Bai&Stone_2011} considered simulations with a vertical domain size $L_z = H$; while appropriate for their unstratified simulations, such a small domain size preferentially captures the damping behavior of ambipolar diffusion that occurs at smaller scales, and would not capture the growing modes of $B_x$ in OA, which are only present at length scales $L_z \gtrsim H$ (Figure \ref{fig:fft_space_comp}). Therefore we anticipate that their empirical criterion may be changed in numerical simulations with a larger vertical domain such as our own, and utilize the linear form of this criterion (Equation \ref{eq:crit_wavelength}) for the remainder of this work.

Although \citet{Bai&Stone_2011} used a suite of simulations to survey the limit of the ambipolar diffusion damped MRI, the magnetic field strength can vary greatly across spatial distances in a single disk. We use the linear criterion of Equation \ref{eq:crit_wavelength} to determine if the ambipolar-damped MRI (i.e. the MRI reduced in strength by ambipolar diffusion) can survive between zonal flows  in which in the strength of the vertical magnetic field can vary.

\begin{figure*}
    \centering
    \includegraphics[width=\textwidth]{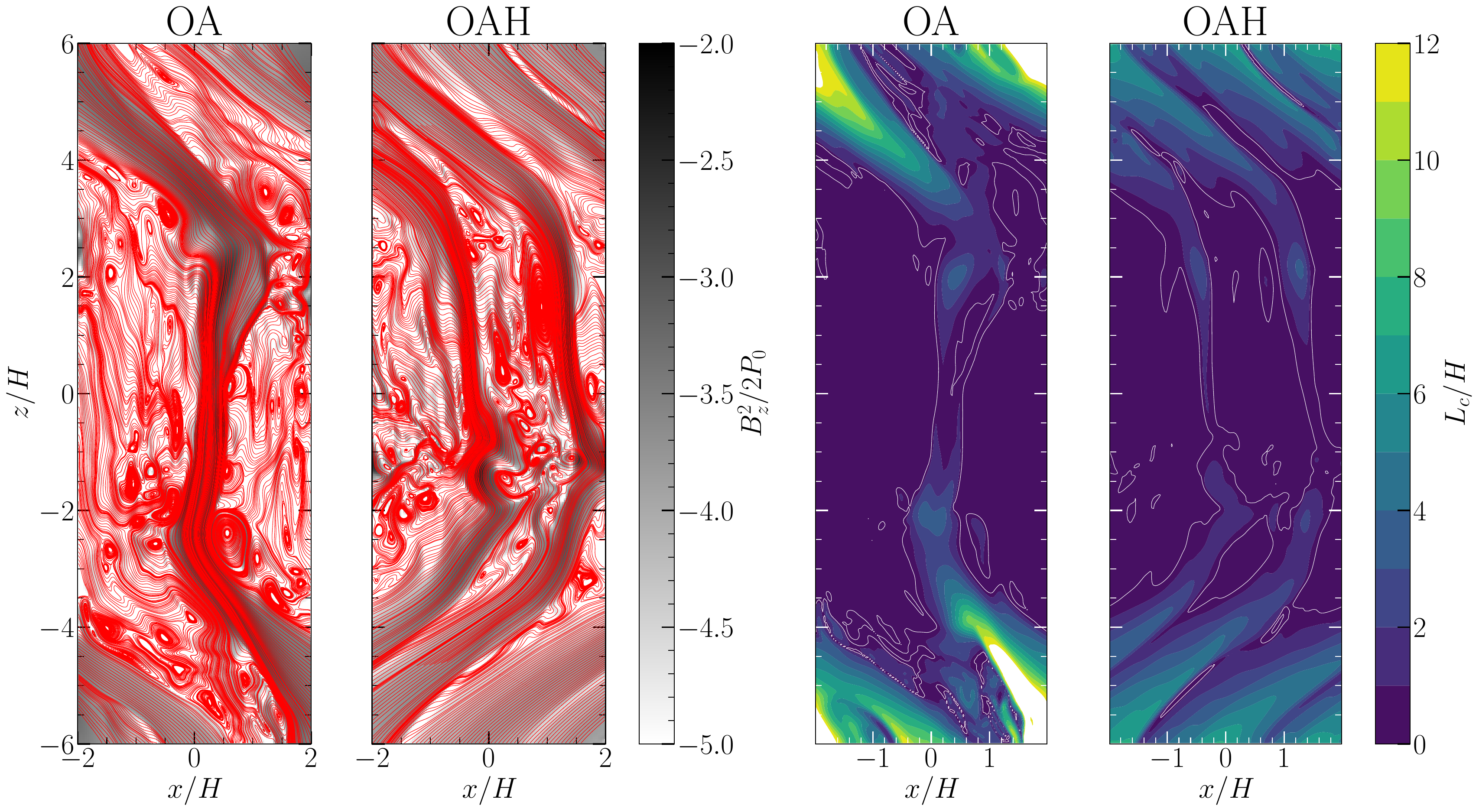}
    \caption{Representative slices of the disk at $y/H = 0$ and at $t~\Omega = 400$, when the magnetic field is radially organized into zonal flows. Left: the vertical magnetic energy (grey scale) with poloidal field lines (red lines). Strong turbulence is anti-correlated with locally strong magnetic field. Right: Contour map of $L_c$, with $L_c/H = 0.5$ (the approximate width of of the vertical field bundles) traced in white. Based on linear theory and assuming that the radial structure can be treated in the same was as the vertical scales (see Section 5 in the main text), darker regions are more likely to contain wavelengths unstable to the ambipolar-damped MRI.}
    \label{fig:xz_slices}
\end{figure*}

Figure \ref{fig:xz_slices} shows a representative slice of the disk for OAH and OA at $t = 400~\Omega^{-1}$, when the magnetic field has had time to saturate.
The left panels of the figure show that the vertical magnetic energy is radially concentrated into bundles of vertical magnetic flux, as is commonly produced via zonal flows in shearing box simulations of non-ideal MHD (e.g. \citealt{Bai&Stone_2014,Riols&Lesur_2019,Rea+_2024}). Red streamlines trace the poloidal magnetic field $\vec{B_p} = B_x\vec{\hat{x}} + B_z\vec{\hat{z}}$. The poloidal field is mostly laminar in areas of large $B_z^2$, and turbulent elsewhere. The right panels of Figure \ref{fig:xz_slices} show the corresponding contour map of $L_c$. Recall that $L_c$ represents the smallest wavelength unstable to the ambipolar diffusion damped MRI; regions of width $L < L_c$ are too small to have the instability present in its linear form. The fact that we see regions of turbulence corresponding to scales larger than the critical scale suggests that the ambipolar-damped MRI is present in those regions.
The vertical field is radially concentrated on scales of $\sim 0.5~H$; the $L_c = 0.5~H$ contour provides a reasonable trace of the radial structure, which appears to reinforce the importance of length scales relative to $L_c$. However, it is not immediately clear why Equation \ref{eq:crit_wavelength}, which was originally derived for a magnetic field with a purely vertical wavenumber, would so readily apply to structure in the radial dimension (radial wavenumber). However, a deeper exploration into the reasons for this are beyond the scope of the current work.

We reiterate that the critical wavelength criterion is derived for the MRI considering only ambipolar damping; without the destabilizing effects of the ADSI. Because the locations in our disk with magnetic fluctuations are well-correlated to the locations with conditions ($\beta$, Am) suitable for the MRI, these results suggest that the MRI is responsible for driving turbulence locally at later times.

Radial concentration of the magnetic field persists at large distances from the midplane. Although the gas approaches ideal MHD close to the disk surface, strong magnetic fields and low densities can cause large \alfven velocities such that

\begin{align} \label{eq:mri_stability}
    (\vec{k}\cdot\vec{v_{\rm A}})^2 > -\frac{d\Omega^2}{d\ln R}
\end{align}

\noindent
which stabilizes the ideal MRI \citep{Balbus&Hawley_1998}. Where the ideal MRI is stable, our simulations instead exhibit a wind-like structure (e.g. \citealt{Bai&Stone_2013_wind}).

\section{Discussion} \label{sec:discussion}

\subsection{Caveats of Linear Theory} \label{sec:discussion:caveats_linear_theory}

Although linear theory is a useful tool in understanding nonlinear, numerical simulations, the reader should be aware of certain caveats to its application.
Firstly, linear theory often assumes the disk is not vertically stratified (e.g. \citealt{Balbus&Terquem_2001,Desch_2004,Kunz&Balbus_2004,Kunz_2008}; but see \citealt{Latter&Kunz_2022} for an example of stratified linear theory).  This neglects the physics of e.g. vertical gravity and buoyancy on linear perturbations. In addition, this assumes that the strength of diffusion is static throughout space and constant in time. The magnetic diffusivities ($\eta_{\rm O},\eta_{\rm A},\eta_{\rm H}$ or their respective Elsasser numbers) are in general expected to vary across space and time due to changes in the strength of external ionization and the magnetic field. 
Thus, while linear theory provides a powerful understanding of the core physics, it is often largely based on assumptions that do not strictly equate to our simulation setup, and one should be cautious in interpreting its connection to numerical simulations.
Besides vertical stratification, there are a number of other assumptions of linear theory. Notably, linear theory often assumes simple exponential time behavior for axisymmetric perturbations to a net vertical field, while in Keplerian rotation the azimuthal field also grows, albeit linearly in time as

\begin{align}
    B_y(t) = B_y(0) - \frac{3}{2}\Omega B_xt
\end{align}

\noindent
To prevent non-exponential time behavior, some studies of linear theory assume that there is no net radial field ($B_x = 0$; e.g. \citealt{Kunz&Balbus_2004}) while other studies place additional constraints on the radial field such that a steady state is achieved with $B_x \ne 0$ (e.g. \citealt{Desch_2004}). This distinction is particularly important to systems that include ambipolar diffusion (and even to non-axisymmetric modes of the ideal MRI; \citealt{Balbus&Hawley_1992}), as $B_y$ explicitly enters into the dispersion relation through geometric terms relevant to the ADSI. Our simulations allow for non-axisymmetric modes. Furthermore, they permit the development of net radial and toroidal fields (due to the vertical outflow boundary conditions); as such, $\langle B_x\rangle_{xy} \ne 0$ and $\langle B_y\rangle_{xy}$ will also grow linearly in time in the presence of shear.

Despite these inconsistencies with numerical simulations, in which perturbations may be non-axisymmetric and the magnetic field may posses more complex behavior in time, small differences in the growth rate predicted by linear theory and the growth rate inferred from simulations are not unreasonable given the limitations in bridging linear theory and more complex numerical simulations. The growth rate prediction $\sigma_{\rm pred} = 0.6~\Omega$ matches extremely well to the observed growth rate of the simulation $\sigma_{\rm sim} = 0.66~\Omega$, which lends support to the usage of linear theory.

\subsection{Numerical Limitations} \label{sec:discussion:numerical_limitations}

The local shearing box approximation carries a number of limitations. These simulations are isothermal, which precludes instabilities that require temperature gradients or more complex cooling prescriptions (e.g. \citealt{Lyra&Umurhan_2019}). Despite these shearing boxes possessing a large vertical domain ($L_z = 12~H$), the shearing box is not a suitable environment for simulating large-scale, magnetically launched winds. For instance, the magnetic field in a global disk must satisfy an ``even-z" geometry with a horizontal current sheet somewhere in the disk, such that material is launched away from the central star on both sides of the midplane (e.g. \citealt{Blandford&Payne_1982}). This is also seen in global simulations, e.g. \cite{Gressel+_2015,Bethune+_2017,Riols+_2020,Cui&Bai_2022,Hu+_2025}, though the sign of the toroidal field does not necessarily change near the disk midplane. The radial symmetry of the shearing box means that it does not have such a restriction. Although these simulations exhibit laminar, vertical (wind-like) outflows at the vertical edges of the computational domain, we refrain from any quantitative analysis pertaining to winds. A thorough discussion on such outflows can be found in e.g. \cite{Fromang+_2013,Lesur+_2013,Bai&Stone_2013_localoutflow}.

\section{Summary} \label{sec:summary}

We have carried out 3D magnetohydrodynamic simulations of a local patch of a protoplanetary disk with and without the Hall effect in order to better understand the origin of the turbulence seen in the simulations of \citet{Rea+_2024} as well as determine the exact role that the Hall effect and ambipolar diffusion play in protoplanetary disk gas dynamics. These simulations possess a stratified ionization profile, which leads to a very large change in the strength of Ohmic diffusion, ambipolar diffusion, and the Hall effect across the height of the simulation domain and are designed to represent the planet forming regions of protoplanetary disks. We find that:

\begin{enumerate}

    \item Three distinct mechanisms are at work in our simulations: the MRI, the ADSI, and the HSI. The ADSI is crucial in driving the initial disk instability, with a mixture of the MRI and ADSI contributing to exponential growth of the magnetic field with a growth rate of $0.66~\Omega$. While the Hall effect measurably enhances existing magnetic fluctuations in the disk midplane, the HSI appears to be of secondary importance in destabilizing the disk.

    \item At later times, the strength of the magnetic field and ambipolar diffusion permit for the ambipolar diffusion damped MRI locally in between zonal flows, which correlates extremely well with the spatial distribution of turbulence.
    
\end{enumerate}

The growth of the MRI and ADSI mixture is extremely fast in our simulations, with magnetic perturbations capable of growing several orders of magnitude in just a few orbits. This short growth timescale suggests that the effects of the ADSI should be seen in situations where the MRI may otherwise have been predicted to be damped by ambipolar diffusion.

These results highlight the importance of a large computational domain in addition to resolving small spatial scales. Unstratified simulations (e.g. \citealt{Bai&Stone_2011}) usually restrict the vertical domain to a single scale height $H$, which also restricts the length scales at which instabilities can operate. Strong ambipolar diffusion in particular will preferentially damp short length scales, which will eventually push the shortest unstable wavelength $L_c$ beyond what can be captured by the computational domain; though the ADSI or MRI may not have much effect if they operate at length scales much larger than the disk itself. In addition to the size of the vertical domain, the vertical stratification itself will also play a role in how the ADSI and MRI operate in the linear regime.

Dust grains will reduce the ionization rate and increase the strength of non-ideal MHD effects \citep{Wardle_2007}. Large grains (Stokes number $\sim 0.1$) are only expected to be in a relatively thin later near the disk midplane, within $\sim 0.1~H$ or so \citep{Lim+_2024,Rea+_2024}. However, smaller grains (Stokes number $\leq 10^{-3}$) can be more significantly mixed with the gas, and may have a particle scale height $\gtrsim 0.7~H$, depending on the strength of gas turbulence; whether turbulence driven by non-ideal MHD effects can survive in such an environment, where the presence of small dust grains may itself feedback onto the system and reduce the ionization rate more, remains poorly understood. 

Overall, these results suggest that while the paradigm has recently shifted toward accretion driven by large-scale laminar winds, turbulence driven by the ADSI and later the MRI can persist in extremely low-ionization environments. In fact, as the initial instability is primarily driven by ambipolar diffusion, a low ionization effect, it may prove difficult for turbulence to be suppressed entirely in protoplanetary disks, and it should not be neglected when considering the mechanisms behind accretion and planetesimal formation.

\section*{Acknowledgments}

D.G.R and J.B.S acknowledge support from the NASA Theoretical and Computational Astrophysics Network (TCAN) program via grant 80NSSC21K0497, the Emerging Worlds program via grant 80NSSC20K0702, and the Future Investigators in NASA Earth and Space Science and Technology program via grant 80NSSC24K1833.
These computations were performed using Pleiades provided by the NASA High-End Computing (HEC) Program through the NASA Advanced Supercomputing (NAS) Division at Ames Research Center.

\software{Athena \citep{Stone+_2008}, NumPy \citep{Harris+_2020}, Matplotlib \citep{Hunter_2007}}

\bibliography{main}
\bibliographystyle{aasjournal}

\end{document}